# An Effective Framework for Managing University Data using a Cloud based Environment

**Kashish Ara Shakil**
Department of Computer Science,
Jamia Millia Islamia
New Delhi, India
Email Id:shakilkashish@yahoo.co.in

**Shuchi Sethi**
Department of Computer Science,
Jamia Millia Islamia
New Delhi, India
Email Id: shuchi.sethi@yahoo.com

**Mansaf Alam**
Department of Computer Science,
Jamia Millia Islamia
New Delhi, India
Email Id: malam2@jmi.ac.in

*Abstract – Management of data in education sector particularly management of data for big universities with several employees, departments and students is a very challenging task. There are also problems such as lack of proper funds and manpower for management of such data in universities. Education sector can easily and effectively take advantage of cloud computing skills for management of data. It can enhance the learning experience as a whole and can add entirely new dimensions to the way in which education is imbibed. Several benefits of Cloud computing such as monetary benefits, environmental benefits and remote data access for management of data such as university database can be used in education sector. Therefore, in this paper we have proposed an effective framework for managing university data using a cloud based environment. We have also proposed cloud data management simulator: a new simulation framework which demonstrates the applicability of cloud in the current education sector. The framework consists of a cloud developed for processing a universities database which consists of staff and students. It has the following features (i) support for modeling cloud computing infrastructure, which includes data centers containing university database ;( ii) a user friendly interface ; (iii) flexibility to switch between the different types of users ; and (iv) virtualized access to cloud data.*

## I. INTRODUCTION

Recently there have been many advances in the technological field. There has also been a growing trend of accessing data and sharing of information such as photographs etc via an internet connection. This has led to development of several new terminologies such as Internet of Things and Cloud Computing. Cloud Computing is a successor of grid computing and is an extension of utility computing where computing capabilities such as servers, network, compute capacities and computing platforms can be accessed as a service via an internet connection. Cloud computing has gradually found its usage as a practice for management of data [16]. It provides an illusion of infinite storage and is tolerant to failures. It is also very robust, highly elastic, cost efficient and scalable [18].Applications of cloud computing are now being used in different fields such as education sector.

There exists a broad gap between the educational opportunities that are available to different learners across the globe. Some learners are privileged and have the best of facilities, infrastructure and instructional materials while others struggle for access to even the most basic instruction facilities [1]. The need of the hour is to develop an effective education system which provides equal educational opportunities to each and every individual. There also exists a large amount of information in educational institutes particular in case of very large universities. This information consists of data about students, teachers, administrative information, student assignments etc. Therefore we need an education system where student's data is managed properly, students have access to instructional materials at all times and existence of a healthy student teacher interaction environment. Furthermore if the instructional material of one university is also made available to students from other universities then this can lead to equal education opportunities to all. In order to meet this end cloud computing comes as a handy solution. Education sector can easily take advantage of the offerings of cloud computing. Cloud computing offers a very cost effective way out for the universities in order to manage their data. Though it has several security issues such as threat to user's data, privacy and others but it is still a popular technique [19]. It helps universities to store their data effectively as it offers infinite storage capacity without any large initial capital investments. Allocation of resources is a potential problem that needs to be addressed in cloud. DMMM [14] and NBDMMM [15] are some of the algorithms that have been proposed for effective allocation of resources in cloud. It enables institutions that lack technical infrastructure and expertise to support a high end computing environment to its researchers on demand as per their requirements. It also offers students, teachers and other academic staff member's access to university data irrespective of their geographical locations.

Many educational institutions have now started shifting towards the cloud by outsourcing e-mail provisions of students. For example Google and Microsoft provide free email services to educational institutes. Some of the educational institutions have hosted their Learning management systems such as a blackboard in cloud. Companies such as Cisco have come up with various cloud solutions for education such as Cisco cloud for schools, Cisco cloud for Higher education and Cisco intelligent network for education [2]. Microsoft's Live@edu is a Microsoft education solution that provides Microsoft services to educational institutes it has now been replaced by Microsoft Office 365.

Both IaaS and PaaS technologies can be used in educational sectors [3]. The use of these technologies in education for management of university data can help the institutions in reaching desired educational goals and objectives. This paper presents a framework for management of educational data in cloud. It also proposes a data management simulator on cloud. This simulator provides features such as support for modeling cloud computing infrastructure i.e. cloud data centers, an interface for users to log on, provision for switching amongst the different users and location independent virtualized access to data at all times.

The remainder of this paper is organized as follows. Section II presents related work done in education sectors; Section III describes the proposed framework. Furthermore section IV presents the proposed simulator finally the paper ends with Conclusion and future works in section V

## II. RELATED WORK

Cloud computing is the hot favorite technology of the year, several companies are now moving towards the cloud. Cloud seems to be the technology buzzword these days. With this growing interest towards cloud even academicians and education sectors are showing a growing affinity towards it. Different cloud based learning platforms are now being used by scholars and universities which provide technical and economical benefits to its users through multimedia based learning materials and innovative strategies for teaching[12].

Moodle and Blackboard are some of the Learning management systems that have begun to be hosted in cloud. The hosting of these learning management systems to a third party helps institution by providing these services at minimal cost and without a huge manpower [5].

In [4] an example of cloud service in education is given. They have explained that how cloud can be used in education through an ePortfolio service. This ePortfolio contains information about students such as their assignments, homework's and academic progress. These ePorfolios can help the respective ministries in improving the entire education system by assessing student's progress and performances.

In [6] the authors have proposed an architecture for management of data in cloud. According to them the management of data in cloud can be divided into three levels: data center level, service provider level and client level. Data center level represents the part where actual storage of data takes place. Cloud service provider level is the middleware level followed by client level. It is at this client level where the education services are offered. The end-users being students, teachers, administrative staff and others.

Cost benefit analysis of cloud computing in education has been presented in [7]. Their work gives a comparison of cost benefit analysis for an organization with 30 users for 3 years. The authors have also discussed a case study of Ben-Gurion University which has shifted to storage on demand. This shift has led to an increase in storage on demand requirements from 65 to 83 percent. It has also demonstrated the capabilities of cloud in providing monetary benefits besides other for the university.

Z Education Cloud [8] is an education cloud based on Mainframes. This education cloud performs functions such as management of user registrations and editing facility of profiles. Course, application and resource management is also taken care of by the education cloud. This cloud is expected to provide rapid deployment capabilities for educational mainframes along with saving of time, manpower and other financial resources.

Many of the authors have further explored the concept of education cloud by extending this concept to a mobile cloud framework where mobile phones can be used for accessing an educational cloud [10]. This form of learning is beneficial for developing and poor countries which lack proper infrastructure such as PCs, bandwidth and other infrastructure facilities. In [11] the authors have devised an intelligent mobile learning education system which is based on a recursive intelligent learning model. This learning model provides personalized leaning options and plans for every individual user.

Knowledge management through cloud has been discussed in [9]. According to the authors knowledge management is a topic of huge concern for both the organizations as well as the educational institutes. They have argued upon the fact that knowledge management particularly in higher educational institutes is emerging as a new field as the institutes no longer have to just provide knowledge but they have to store this knowledge for future references. The proposed framework in this paper also attempts to solve this knowledge management issue that occurs in higher education institutes particularly in case of very large universities where the number of students is huge.

Therefore, even though cloud computing has found its way as an effective technique in the education system but the way in which this technique is accepted amongst the education fraternity is equally important. Thus, there is a need for knowing the extent to which the students use the education cloud. In [13] a study of usage of cloud for education is done. As per this study the student usage of cloud is maximum for communication and minimum for cloud documents and there exists a strong correlation between the need for cloud and its reported usage.

# An Effective framework for managing university data using a cloud based environment

## III. UNIVERSITY DATA MANAGEMENT FRAMEWORK

Higher educational institutes are overburdened with huge volumes of data and there is a need for management of data involved in such educational institutes in order to improve the overall quality of education which is imparted in such institutes. The proposed framework is a framework for management of data in such higher education institutes. The proposed architectural framework is inspired from the three schema cloud data base management system architecture [6].According to the proposed architecture for management of data in universities shown in Fig. 1, there exist three levels at which management of data takes place in large universities: data center level, service provider level and the educational users or academic user level. The data center level consists of actual physical data storage, cloud service provider level is managed by the service providers for providing services to cloud end users and the end user level consists of cloud end users which usually include educational cloud end users.

### A. Data Center/Physical Storage Level

The data center level usually consists of the actual physical storage location where the entire data is stored. In case of education cloud this data usually consists of student information, student assignments, teaching materials and online lectures etc. This level is also responsible for maintaining several servers offering virtualized storage capabilities to its end users. This storage is provided in an elastic and scalable manner on a pay per usage manner thereby exploiting the advantages offered by cloud.

### B. Cloud Service Provider Level

The cloud service provider level usually involves the role of a cloud database administrator. This administrator is responsible for making sure that all the services requested by the clients is provided to them on fly without any issues. This level provides the end users of an educational cloud with the illusion of infinite storage capacity, network bandwidth and compute capacity. It also offers various software's that are required by the university cloud users as a service thereby helping the users to make use of these software for learning purposes without any hassles in installing and purchasing expensive software's. It helps in inculcating an effective learning environment amongst its users where the learners can just concentrate on their core tasks i.e. to learn and develop newer skills. This level is also responsible for ensuring the availability of cloud services at all times. Administrative tasks such as availability of services and user accounts are handled at this level along with maintenance of privacy of information about users and handling of security issues and potential threats to security.

### C. Educational User Level

The educational user level generally consists of the end users of an educational cloud .At this level an end user is exposed to only those services that they have requested. For example if a learner has requested usage of software such as matlab then only matlab will be made available to him or her. The end users in an education cloud usually consist of students, research personnel's, academic staffs and other administrative staff. The users might access the university cloud at this level by using an API provided to them via the service providers.Thus the proposed architectural framework aims to provide a solution for the management of data in large universities where storage, retrieval, processing and management of information is a challenging task. The following are some of the features of the proposed framework:

- Support for modeling cloud computing infrastructure, which includes data centers containing university database
- A user friendly interface
- Flexibility to switch between the different types of users
- Virtualized access to cloud data.

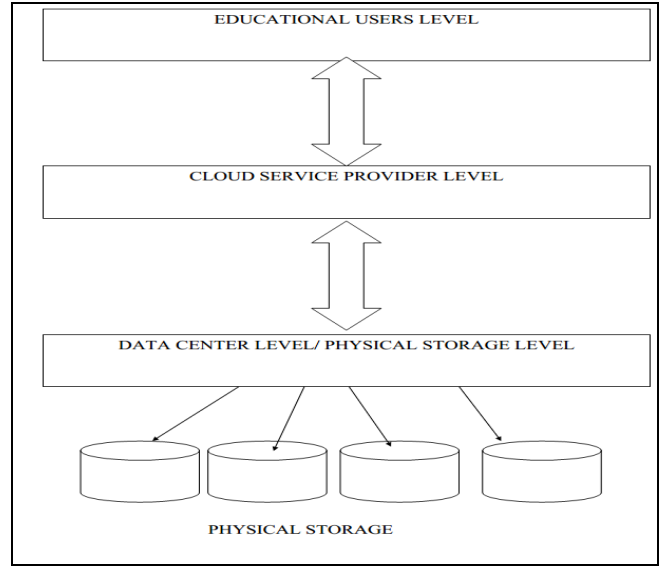

Fig. 1. University Data Management Framework

## III. UNIVERSITY CLOUD SIMULATOR

In order to demonstrate the applicability of the proposed framework a simulator was developed on java platform and using oracle 10g. Table I shows the software and hardware details and requirements for the simulator. This simulator provides a demonstration of working of a university cloud.

### A. Case Study: Jamia Millia Islamia

This simulator has been developed for Jamia Millia Islamia which is a central university located in New Delhi, India. Being a central university in India there are lakhs of students enrolled. Maintenance of information about every student, collection of assignments, submission of assignments and updating of information are cumbersome tasks especially for such a huge number of students. Therefore management of all this data has always been a cause of concern for the university. Though the university has been making use of technologies such as management information system for managing student

TABLE I
HARDWARE AND SOFTWARE REQUIREMENTS FOR SIMULATOR

| Hardware/Software Requirements | Details |
| --- | --- |
| Processor | 233 MHZ or above |
| RAM Capacity | 256MB |
| Hard Disk | 20GB |
| Operating System | Windows NT/2000 |
| Services | JDBC |
| Database | Oracle 10g |

Information but it still fails to provide student friendly services such as interface for students to access their information, uploading of assignments, access to course study material etc.

Therefore, there is a need for the universities to develop such a system which is beneficial for every individual affiliated to the university and even for people who are outsiders. It is here that cloud technologies come into picture where management of student information can be done easily. The University can use a cloud setup which provides services to its users for proper access and management. Thus, the university cloud simulator has been developed taking all the requirements into consideration. This simulator has four modules: terminal, cloud, student and staff module.

1. Terminal Module: This module is the interface of the university cloud. It offers facilities such as desktop interface, access to virtual machines and software as a service to the users. It is through this interface that the staff and students can log on to the university cloud. The entire input and output operations take place via this module. The input operations usually consist of entering student information, retrieving information, upload of assignments etc.

2. Cloud Module: This module simulates the service provider functions of an actual cloud setup. This module is responsible for catering to the needs and requirements of staff and students. All the services demanded by the end users is handled at this module.

3. Student Module: This module is responsible for processing all the student requests. It is also responsible for accessing, retrieving and updating student information

4. Staff Module :This module is responsible for processing all requests of staff members. It handles all the functions available to the staff users in the university cloud.

Figure 2 shows the snapshots of home interface of the university cloud simulator. Figure 3 shows the general interface of the module. Figure 4 shows the snapshot of the simulator when a student performs insert operation in the simulator. Figure 5 shows the snapshot of the simulator when a student performs the retrieve operation in the simulator.

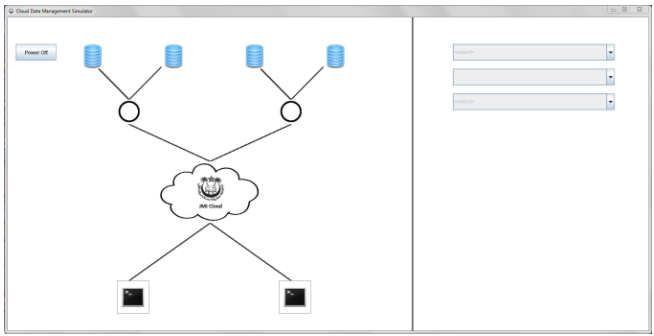

Fig. 2. Snapshot of Home Screen of simulator

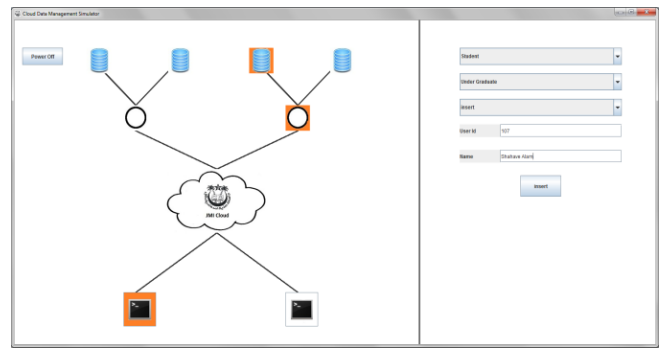

Fig. 4. Snapshot of Student Insert Operation

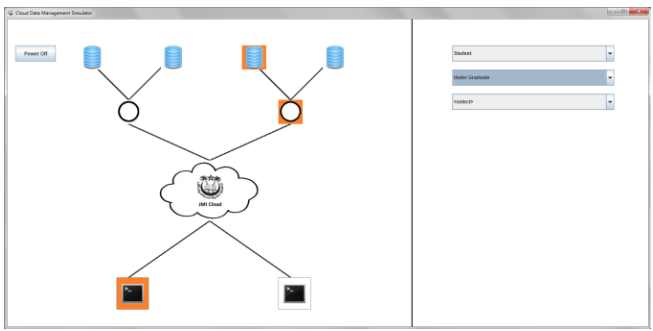

Fig. 3. Snapshot of Intermediate Screen of simulator

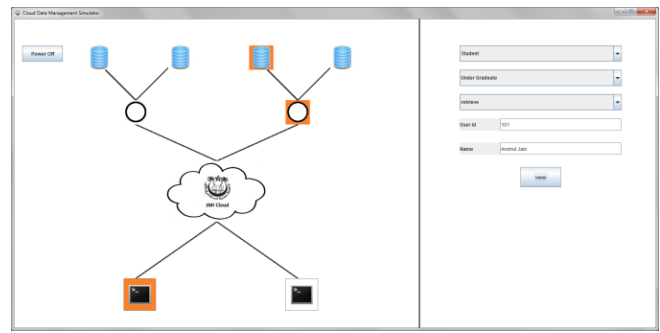

Fig. 5. Snapshot of Student Retrieve Operation

An Effective framework for managing university data using a cloud based environment

| Test Case | Check Item | Test case Objective | Steps to Execute | Test Data / Input | Expected Result |
|---|---|---|---|---|---|
| TC-01 | Duplicate user Id | Data with duplicate user id | Click Insert | Duplicate User Id | Data must not be inserted |

Fig. 6. Test CaseDesign for Inserting Data

| Test Case | Check Item | Test case Objective | Steps to Execute | Test Data / Input | Expected Result |
|---|---|---|---|---|---|
| TC-02 | Wrong user id | Check for wrong user id | Click view | Wrong user id | No user found. |

Fig. 7. Test case Design for Retrieving Data

B. Testing of the Simulator

A thorough testing of the simulator was done, beta testing was done for the simulator and then changes were made after the errors were reported.

Apart from this testing of the individual operations were done, using the test case design given by Fig. 6 and Fig. 7.

## IV. CONCLUSION AND FUTURE SCOPE

Cloud Computing is the latest technology and has found widespread usage and applications. Many people from different domains have found applications and usage of cloud in their respective fields. Even the education sector has acknowledged the widespread usage of cloud and thus, there has been a growing inclination of education sector towards cloud.Many learning systems have now started being hosted in cloud. This paper is one such attempt to show the applicability and usage of cloud in education sector. The usage of cloud in education sector provides several benefits such as economic benefits, location independent access to data, enhanced learning experience and increased interaction amongst the educational community worldwide. In this paper an architectural framework for management of university data is proposed.Furthermore, a simulator for a central university in India is developed on java platform using oracle 10g. It has several features such as support for modeling cloud computing infrastructure, which includes data centers containing university database,a user friendly interface and flexibility to switch between the different types of users.

For the future it is planned to implement this simulator using hadoop (HDFS) instead of oracle 10g. It is also planned to extend the capabilities of this simulator for handling big data.

ACKNOWLEDGEMENT

The authors would like to thank Jamia Millia Islamia for funding the project under Innovative Research Activities grant.